\documentstyle[seceq,epsf,psfig]{ptptex}

\def\DM{Dzyaloshinsky-Moriya }

\title{
Static and Dynamical Dzyaloshinsky-Moriya interactions in gapped spin systems

}

\author{
Olivier {\sc C\'epas}$^{a,b}$,  and  Timothy {\sc Ziman}$^{a}$\footnote{Also at the CNRS, LPM2C, UMR 5493, Grenoble} \footnote{E-mail addresses: cepas@physics.uq.edu.au,ziman@ill.fr}
}

\inst{
$^a$Institut Laue Langevin, B.P. 156, 38042 Grenoble, France. 
\\
$^b$ Department of Physics, University of Queensland, QLD 4072 Australia.

}

\abst{

Anisotropic spin-spin interactions of  Dzyaloshinsky and Moriya  symmetry
are generally considered weak, as they depend
on the spin-orbit couplings. In spin systems with gapped
ground states they can, however, have rather strong effects. We  will discuss
recent results related to the  results of neutron scattering and ESR 
for  $\rm SrCu_2(BO_3)_2$.
Inclusion of the Dzyaloshinsky-Moriya interactions can
explain much of the  dynamics of the systems discussed, in particular the 
splitting and dispersion of the triplet modes. Some
effects  remain to be explained, however. 
Symmetries of the crystal  lead to the prediction of zero intensity 
for transitions, for example between the ground state and the
triplets  observed in ESR.
 
We present
recent calculations of the effects of  anisotropic terms  
generated dynamically,
ie linearly in the phonon coordinates.  We discuss how
this leads to a novel mechanism to explain the ESR intensities. 
For polarized  neutron scattering
experiments, we can calculate the mixing of nuclear and
magnetic scattering amplitudes by such a term and how such mixing should be observed.

} 
\begin{document}

\maketitle

\section{Introduction}
In  quantum spin $\frac{1}{2}$ magnets with  
the leading anisotropies in powers of the spin-orbit coupling are
\DM interactions\cite{rf:Dzyaloshinski,rf:Moriya} $\sum_{i,j}  \vec{D}_{i,j}
.(\vec{S}_{i} \times \vec{S}_{j})$, (with sum over  neighbours $i$ and $j$),
which appear in first
order  and anisotropy in the 
exchange, which appears in second order. In the copper oxides such terms 
are an order of magnitude  weaker than the isotropic exchange  $D = (\frac{\Delta g}{g}) J$ with $\Delta g = g-2$ yet we will see that they
can have strong effects on  the dynamical response. 
In a gapped system with singlet ground states, 
unlike an ordered antiferromagnet, the relatively weak 
\DM terms will not close the gap to allow, for example, 
weak ferromagnetism with  a spontaneously broken symmetry. 
The interactions can have important consequences in addition to 
an explanation
of  transitions otherwise forbidden by overall spin rotation symmetry. 
They may give splittings {\it linear} in the spin-orbit strength
which cannot be cancelled by the higher order terms in the 
exchange\cite{rf:Aharony,rf:Maekawa}.
They may  also
enlarge the effective magnetic unit cell,
giving extra branches to excitations.
Furthermore  the dispersion of excitations due to the stronger isotropic
exchanges may be strongly reduced by frustration, allowing the 
splittings of the  \DM terms to dominate. 
We will illustrate these points by referring to recents experiments and theory
in the  geometrically  frustrated $\rm SrCu_2(BO_3)_2$.

While the \DM 
interactions lower the spin symmetry allowing certain transitions
forbidden from  a completely isotropic singlet state,
it turns out that  there are transitions observed that should be forbidden
even in their presence. 
This  leads us to consider a higher order
of anisotropy: ``dynamical \DM'': spin-phonon terms in the Hamiltonian
which modulate not only the exchange strength but in addition modify the anisotropic
terms of \DM form. This is especially 
interesting when the  terms  generated
are forbidden in the equilibrium structure.
We can treat such generalised spin-phonon terms perturbatively
and for each operator corresponding to a physical process
generate an effective operator
in terms
of spin-operators.  These
may explain optical transitions, at wave vectors $q = 0$, observed by ESR and infrared absorption,
and, for finite values of $q$, mixing of nuclear and magnetic neutron scattering amplitudes. 
\par

\par
This paper will review material presented in greater detail either for the 
static \DM  \cite{rf:borate,rf:SakaiCZ} and the dynamic  \cite{rf:these,rf:OlTim}.

\section{Dynamics: Examples of the influence of \DM in $\rm SrCu_2(BO_3)_2$}
{$\rm SrCu_2(BO_3)_2$} is  exciting in that 
it can be considered 
as planes of spins $\frac{1}{2}$ interacting via 
the Hamiltonian of the Shastry-Sutherland model in two dimensions.
The  interaction between planes
is via couplings that are both weak and frustrated. 
The Shastry-Sutherland  model
has the peculiarity that the product of singlet
states on the closest dimers with the stronger exchange $J$ is still an {\it exact} eigenvector when the frustrated second nearest neighbour interactions $J^{\prime}$ 
are included\cite{rf:ShastryS}. Furthermore this eigenvector is the ground state 
even for the relatively large value
of the relative coupling $J^{\prime}/J = 0.62$. 
While the ground state does not change they dynamics are strongly renormalized
by the weaker coupling, and in fact the 
coupling is close to the estimated value  $J^{\prime}/J \approx  0.68$
where there is a quantum phase transition, possibly to  a plaquette
state\cite{rf:Koga2000}. 
This ratio is estimated either from the susceptibility\cite{rf:Miyahara} 
or the 
ratio of the energies of singlet states, seen in Raman scattering,
to triplet energies, seen by magnetic neutron scattering\cite{rf:borate}.
\subsection{Splitting of the triplet mode by the  \DM interaction}
It was observed in \cite{rf:Nojiri2}, however, that the lowest triplet state is
not simply renormalized in energy by $J^{\prime}$
but split as well, and this was confirmed by neutron scattering
which gave the dispersion. This lead us to  consider the \DM
interactions 
in $\rm SrCu_2(BO_3)_2$ from the usual symmetry rules of 
Dzaloshinsky and Moriya.
Apart from a small ``buckling'' of the planes ( which we shall mention later)
the presence of a centre of inversion  leads to zero \DM interaction between spins of the more strongly coupled  dimers,
and for the more weakly interacting dimers, as the  plane $(ab)$ of the spins 
is a mirror plane,  \DM vectors are strictly perpendicular  with alternate signs as shown in 
Figure 1, taken from reference\cite{rf:borate}.
\begin{figure}[htbp]
\begin{center}
\leavevmode
\psfig{file=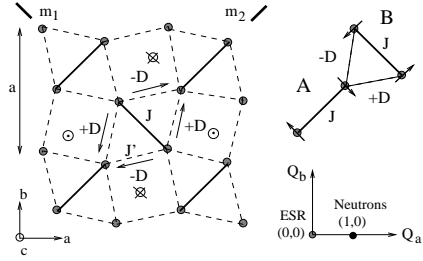,width=5.5cm,angle=0}
\end{center}
\caption{\DM vectors determined from the symmetry, ignoring the 
weak buckling of the (ab) plane. The arrows indicate the sense
of the neighbours $i$ and $j$ in the definition of the vectors which are
perpendicular to the plane with directions indicated} 
\end{figure}
Inclusion of such an anisotropy will  give corrections to the 
Shastry-Sutherland ground state but these can be estimated perturbatively.
To understand the effect of the dynamics one can first ignore $J^{\prime}$
and calculate the dispersion coming from the term $D$, which gives
a dispersion even in linear order, corresponding physically to the 
fact the the \DM vectors are not frustrated:
The dispersion of
the two modes $\pm$ (each is twice degenerate with $S^z=\pm 1$) is
therefore proportional to $D$ :

\begin{equation}
\omega_{\bf{q}}^{S^z=\pm 1,\pm} = J \pm 2D \cos(q_a a/2) \cos(q_b a/2) 
\label{gap}
\end{equation}

\noindent 
where $f({\bf{q}})= \cos(q_a a/2) \cos(q_b a/2) $. 
On the other hand, the Dzyaloshinski-Moriya interaction has no effect on the $S^z=0$
component of the triplet, so that its energy remains equal to $J$ ($\omega_{{\bf{q}}}^{S^z=0}=J$) (fig. \ref{resume}). 
Thus for $J^{\prime} = 0$ the two transverse modes will be
split by $4D$. Physically the splitting into two modes
results from the fact that without the \DM vectors the effective
magnetic unit cell is two times smaller, as the two dimers
per unit cell are equivalent, at least in the limit
of small  $J^{\prime}$. The different signs of the \DM vector
to the left and right of a given dimer lower the effective symmetry,
doubling the number of modes.  As $J^{\prime}/J$ is not small, the numerical factor
of 4 is renormalised to be a function of  $J^{\prime}/J$ and this we have calculated by finite size scaling
on small clusters, as shown in Figures 2 and 3. Even close to the quantum phase transition the gap to the triplet while renormalised to smaller values
remains large and such calculations converge quite rapidly.
Using the experimental value of  $J^{\prime}/J$ the factor is renormalized to
2.0 \cite{rf:borate}. From the experimental values from   
the optical experiments of Nojiri et al\cite{rf:Nojiri2} and the neutron
inelastic scattering\cite{rf:borate,rf:KakuraiFuku}, we can deduce the absolute value of 
the \DM  interaction $\vec{D}^c = 0.18 $ meV. 
\begin{figure}[htbp]
\centering
\parbox{5.5cm}{
\psfig{file=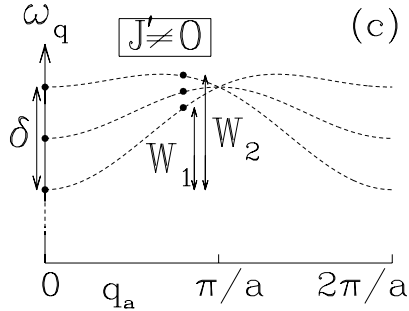,width=5.5cm,angle=-90}
}
\vspace{0.2cm}
\caption{For finite $J^{\prime}$,  exact diagonalization for a cluster of 
20 spins gives the energies of two reciprocal points (the dots)}
\end{figure}

\begin{figure}[htbp]
\begin{center}
\leavevmode
\psfig{file=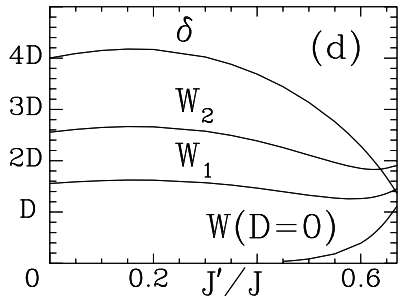,width=5.5cm,angle=-90}
\end{center}
\caption{Renormalization of the splitting $\delta$ and the widths $W_1$ and $W_2$ defined from Fig 2. $W$ is the bandwidth
for the dispersion for zero $D$,  with   scale corresponding to the physical value of $D$.}
\label{resume}
\end{figure}

\par
The same cluster calculation can be used to estimate the dispersion (see the lower line in Figure 3) which is relatively weak as it begins at sixth order
in  $J^{\prime}/J$.  
It is striking that  this splitting
can dominate the dispersion even though it is determined by an interaction
an order of magnitue smaller than the isotropic interaction.
\par
While this theory gives good quantitative explanation of the optical
and neutron experiments with wave vector
the most recent high resolution experiments\cite{rf:KakuraiFuku}
indicate possible discrepancies near $(q_a,q_b)=(\pi,0)$. The gap at that point is
 interpreted as a sizeable
transverse component to the \DM vector. This had been assumed to be 
negligible  as the buckling in the plane
responsible for breaking the symmetry that would forbid it is small.
We must still add the 
proviso that  while it appears now that a transverse component should
be included, especially in view of the behaviour in finite magnetic field\cite{rf:KakuraiFuku}, a  reliable quantitative estimate is not yet
possible, especially as the gap is smaller, close to the estimated bandwidth from the 
 $J^{\prime}$ terms. A reliable calculation will probably need diagonalisation of larger clusters
\cite{rf:Pakwo}.

\subsection{Selection rules for \DM interaction}
In the optical experiments of Nojiri et al\cite{rf:Nojiri2}, the resonance is from the 
ground state to the excited magnetic states. The observation of absorption
requires some anisotropies: as the ground state without anisotropies
is a spin singlet the operator corresponding to  coupling with the probe magnetic
field $\vec{h}.\sum_{i} S_i$ applied to the ground state  vanishes. As the \DM interaction
mixes in non-singlet components the matrix elements to  excited states
may be non-zero. We must use the  symmetries to predict which ones are non-zero
\cite{rf:SakaiCZ}.
For  {$\rm SrCu_2(BO_3)_2$ }, a lattice symmetry (reflection in a diagonal
followed by rotation by $\pi$) leads to a zero amplitude for excitation
of the triplet states, even in the presence of the 
\DM couplings. As we have mentioned for the dispersion,  there are additional
anisotropies due to buckling of the planes and anisotropies of the 
$g $ tensors, but nevertheless the amplitude of the absorption 
in the two cases is somewhat surprising, and this leads us to consider an
alternative explanation in terms of a {\it dynamical} \DM interaction. 
\section{Dynamical \DM interaction }
We shall now consider a general anisotropic
spin-phonon couplings corresponding to modulation of the 
exchange by linear coupling to lattice distortions. The 
term in the Hamiltonian coupling the phonon and spin operators is:

\begin{equation}
{\cal H}^{\prime} = \sum_{ijd\alpha\beta} g_{d}^{\alpha} u^{\alpha}_{id}
\vec{S}_i.\vec{S}_{j} + d^{\alpha \beta}_d u^{\alpha}_{id}
(\vec{S}_i \times \vec{S}_{j+1})^{\beta} 
\label{spin-phonon coupling}
\end{equation}
\noindent 
where $u^{\alpha}_{id}$ is the $\alpha$ component of the displacement
operator of atom $d$ in unit cell $i$, $g_d^{\alpha}$ and $d^{\alpha
\beta}_d$
are, respectively, the isotropic spin-phonon coupling and the
dynamical Dzyaloshinsky-Moriya interaction.
For example the inversion symmetry that forbids the \DM interaction
within the strongly bound dimer in  {$\rm SrCu_2(BO_3)_2$
may be lifted instantaneously by a phonon mode, generating a 
\DM anisotropy.
Consideration of ``dynamical'' \DM terms were in fact motivated
first by experiments in inelastic neutron scattering.

\subsection{Polarisation analysis of inelastic neutron scattering}
In another paper in this symposium\cite{rf:Regnault_Fukuoka}
L.P. Regnault had discussed how  experiments with  polarisation
analysis in all directions of incident and scattered beams of neutrons
can give new information on correlation functions. One can probe mixed ``nuclear'',
and ``magnetic'' correlations\cite{rf:Blume,rf:Maleyev_neut}.
If  a rotation of the outcoming
spin polarisation around some axis can 
be measured, only an interaction with a ``handedness'' such as the \DM
interaction can give a non-zero result\cite{rf:Maleev_pol}.  
  Such experiments are therefore ideal for their sensitivity to 
the dynamical \DM terms we have discussed here. 
Here we will give a simple introduction to the contribution
of the  terms to inelastic scattering.

Let us consider an incident beam of neutrons fully polarised, let us 
say in the ``up'' direction of an axis $z$. The spin part of the 
incident wave function
is then 
\begin{equation}
\psi_i = |  \uparrow >
\end{equation}
\noindent
The neutron is scattered both by the atomic nuclei via the strong interaction  and the  coupling via its spin to the magnetic field generated
by the  magnetization of the sample. The spin part of the outgoing wave function 
for momentum transfer $\kappa$  is, by the Born approximation,
\begin{eqnarray}
\psi_o &=& ({\cal N} + {\vec{\cal M}}^{\perp}_{z})| \uparrow > + ({\vec{\cal M}}^{\perp}_{x}+i{\vec{\cal M}}^{\perp}_{y} |  \downarrow > \\
 {\cal N}(\kappa) &=& <\phi_{\alpha}|\sum_{id}a_d\exp{i\vec{\kappa}\cdot {\vec r}_{id}} | \phi_{0}> \\ 
{\vec {\cal M}}^{\perp}(\kappa ) &=& <\phi_{\alpha}| {\vec{\kappa}} \times (\vec{S}_{\vec{\kappa}} \times \vec{\kappa} ) | \phi_{0}>  
\end{eqnarray}
\noindent
Here $\phi_{0}$ is the ground state and $\phi_{\alpha}$ the  state
excited by inelastic scattering of the neutron. ${\vec {S}}_{\vec{\kappa}} $ is the Fourier transform of 
both spin and orbital contributions to the magnetization.
The momentum transfer $\kappa$ can be chosen by the geometry of the 
experiment:  it is frequently  taken parallel to the axis of polarisation in order
to eliminate the term ${\vec{\cal M}}^{\perp}_{z}$. In this configuration
the spin of the scattered neutron will then be simply $\sigma_z = (|{\cal N}|^2-|{\vec{\cal M}}|^2)/2$
In this case the ``nuclear'' term $\cal N$, 
is simply separated from the 
``magnetic'' $\cal M$ by the amplitude of non-spin-flip to spin-flip
scattering. When we resolve in energy, if there is separation
of the nuclear and magnetic degrees of freedom in the Hamiltonian
either  $\cal M$ or $\cal N$ vanishes and the coherent scatted beam will be either
fully polarised ``up'' or ``down''. If the excitations are mixed, then
it becomes interesting to measure, for example, the spin in another direction.
We shall consider the expectation value of the spin in a transverse
direction $x$. From the Pauli matrices this is just  $\sigma_x =\left[ ({\vec{\cal M}}_x+i{\vec{\cal M}}_y)^* {\cal N}+{\cal N}^* ({\vec{\cal M}}_x+i{\vec{\cal M}}_y)\right] /2$. Thus by measuring in a transverse direction we
have an alternative measure of the two amplitudes. It  has the advantage
that if one is small, the result is linear rather than quadratic and 
that by measuring in two perpendicular directions we can also
measure relative phases of  $\cal M$ or $\cal N$.  
Interpretation of the polarisation requires calculation
of the  matrix elements $\vec{\cal M}$ and $\cal N$ in a case when they
are both non-vanishing. 
Here we shall consider the case of $\cal N$,
the nuclear scattering amplitude, for an excitation considered ``magnetic'':
ie which in the absence of spin-phonon terms would have vanishing
nuclear amplitude.
The  matrix element of the nuclear operator   between the perturbed states $0^{\prime}$ and $\alpha{\prime}$ including $\cal{H}^{\prime}$ can then be written as that of an effective
operator acting between the unperturbed states $0^{}$ and $\alpha{}$. This
operator is purely written in terms of spin operators:

\begin{equation}
\langle \alpha^{\prime} | \sum_{id} a_d \exp{i{\vec{\kappa}\cdot \vec{r}_{id}}} |    0^{\prime} \rangle = 
\langle \alpha \mid \sum_{ij} \exp ({i\vec{\kappa}\cdot \vec{R}_i}) \left( \gamma_{\vec \kappa} \vec{S}_i . \vec{S}_{j} + \vec{\delta}_{\vec \kappa}.(\vec{S}_i \times \vec{S}_{j})\right) \mid 0 \rangle  
\end{equation}
\noindent
where  $\vec{R}_i$ is the equilibrium position of the atom $i$ and  the two terms depend on  $\gamma_{\vec \kappa}$ and $\vec{\delta}_{\vec \kappa}$, 

\begin{eqnarray}
\gamma_{\vec \kappa} = i \sum_s \frac{\Omega_{\kappa s} a_s(\kappa) }{\Omega_{\kappa s}^2 - \omega_{\kappa}^2} g_{\vec{\kappa}s} \label{notation0} \\
\vec{\delta}_{\vec \kappa} = i \sum_s \frac{\Omega_{\kappa s} a_s(\kappa) }{\Omega_{\kappa s}^2 - \omega_{\kappa}^2} \vec{d}_{\vec{\kappa}s} 
\label{notation}
\end{eqnarray}
\noindent 

Note that $\vec{r}_{id}$ is the (instantaneous) position of an atom, i.e. with
phonon mouvement included, in contrast to $\vec{R}_i$ (where we have suppressed the factor $d$ to indicate that we take terms including only one
magnetic ion per unit cell). Thus we have integrated out the phonon motion
by perturbing the excited and ground states to first order in the spin
phonon coupling.
We  now need only calculate matrix elements of an effective spin Hamiltonian.
The coefficients are defined as follows:
 ${a}_s(\vec{\kappa} ) = \sum_{id} a_d \exp{i{\vec{\kappa}\cdot \vec{r}_{id}}}$ 
is the nuclear form factor
of  the  phonon
mode $s$.
Its magnitude is measurable independently from the intensity 
of {\it real} phonon
scattering at  energy $\Omega_{\kappa s}=\Omega_{({\vec q}={\vec \kappa},s)}$.
The final magnetic
state has an energy $\omega_{\alpha}$. $g_s=\sum_{d,\alpha} g_d^{\alpha}
\lambda^{\alpha}_{\vec{\kappa}ds}$ is the amplitude of the variation of the
magnetic exchange energy due the atomic distortions of the phonon $s$
($\lambda^{\alpha}_{\vec{\kappa}ds}$ is the amplitude of the motion of the atom
$d$, in the direction $\alpha$ due to the phonon $s$ at $q=\vec{\kappa}$).
Here the sum $ij$ is assumed to run over a set of equivalent neighbours:
more generally there could be a set of $\gamma$ and $\delta$ for different
inequivalent neighbours.
A particular phonon mode $s$
contributes only if 
${a}_s(\vec{\kappa}) \neq 0$: the virtual phonon $s$ creates distortions that have non-zero nuclear amplitude. That is, the phonon is not purely transverse.
The  two terms may then be generated provided:
 \begin{itemize}
\item $g_{\vec{\kappa}s} \neq 0$: The distortion of the unit cell due to the phonon
$s$ modulates the magnetic exchange between the spins. This term
can give non-spin-flip transitions to excited singlet states
at zero temperature:
$\Delta S_{tot}=0$ are allowed.
\item $\vec{d}_{\vec{\kappa}s} \neq 0$:  The distortion of the unit cell 
due to the phonon $s$ must break instantaneously the symmetry by inversion 
at the middle of the bond; so as to allow an 
instantaneous \DM interaction of amplitude $\vec{d}_{\vec{\kappa}s}$. 
Directions of the vector  $\vec{d}_{\vec{\kappa}s}$ 
are constrained by the symmetry rules for static \DM
interactions applied to  the equilibrium structure distorted by the given phonon
${s,\vec{\kappa}}$. 
Transitions between different spin states $\Delta S_{tot} =0,\pm 1$ are allowed.
\end{itemize}
In the second case we see that scattering to the triplet state, normally
purely spin-flip will have a small component transverse from the 
nuclear amplitude. There is then a rotation away from the pure spin-flip direction of the spin of the scattered neutrons by an angle  that is
essentially $|\frac{\cal N}{\cal M}|$.
This gives an estimate of the rotation of the polarisation: 
\begin{equation}
 \left( \frac{\Delta g}{g} \right) \left( \frac{\Omega_{\kappa s} E}{\Omega_{\kappa s}^2 - \omega_{\alpha}^2} \right)
\sqrt{\frac{\frac{d^2 \sigma}{d \omega d\Omega}_{phonon\  s}}{\frac{d^2 \sigma}{d \omega d\Omega}_{triplet\ \alpha}}}
\label{ordre de grandeur}
\end{equation}
\noindent
The factor $E$ here is the modulation of the isotropic exchange
and can be approximately estimated from the contribution of the phonon to 
change in the angle of superexchange. 
The factor $\frac{\Delta g}{g}$ is from the usual Moriya estimate of the 
anisotropic part. The ratio of inelastic cross sections $(\frac{d^2 \sigma}{d \omega d\Omega})$
is , as already mentioned,  measurable independently from the relative 
intensities of real phonon emission at the phonon frequency to the  (spin-flip) magnetic scattering.
The angle about which the polarisation will turn depends on the 
vectors  ${\vec d}_{s,\vec{\kappa}}$. 
Estimates of the 
values expected for  rotations\cite{rf:these,rf:OlTim} expected in copper oxides give results that  are  a few degrees in the most favourable cases, and in general
much smaller, essentially because of the double constraint on both
the form factor of the phonon and the generation of spin anisotropy.

We have performed a similar calculation to the above for 
the possibility of electric field induced optical transitions\cite{rf:these,rf:OlTim} We do not enter into details here
but note that
in calculating the relevant matrix element for electric dipole absorption
to a magnetic state,  while
the same magneto-elastic constants and vectors will enter
and there is an effective matrix element of the same form as at $\kappa = 0$
but the  vectors $\vec{\delta}$
will differ as ${a}_s(\kappa)$, for example, will be replaced
by an optical  form factor.

\section{Conclusions}
We have reviewed effects of both static 
\DM interactions and terms generated by coupling to phonons
that lower the symmetry in the spin gapped compound {$\rm SrCu_2(BO_3)_2$.
While the observed magnetic modes are well explained by inclusion of 
\DM interactions, some puzzles remain. We have advanced the idea that
some of these puzzles may be resolved by 
the dynamic terms. Such terms have the effect of mixing  nuclear and magnetic
scattering amplitudes  in neutron scattering, and allowing excitation
by the electric field component of the probe electromagnetic
field to excited  magnetic states.
Future  neutron experiments
with full  analysis of the spins and 
optical experiments  using  
polarisation of the electric and magnetic components of the light
should test these ideas.

\section*{Acknowledgements}
We would like to thank
T. Sakai for continuing theoretical collaboration. We worked closely with the experimental teams
of J-P Boucher, L-P Regnault, and K. Kakurai and collaborators, and benefited
 from results generously communicated by H. Nojiri.

\appendix


\end{document}